\documentclass[twocolumn,showpacs,amsmath,prb,superscriptaddress]{revtex4-1}
\usepackage{epsfig,psfrag}
 \usepackage{amsbsy}
\usepackage{amsfonts}
\usepackage{wasysym}
\usepackage{bbm}
\usepackage{tabularx}
\usepackage{euscript}
\usepackage{amsfonts}
\usepackage{exscale}
 \usepackage{slashed}
\usepackage{subfigure}
\usepackage{comment}

\usepackage{amsmath,amssymb,wasysym}
\usepackage{graphicx}
\usepackage{dcolumn}
\usepackage{bm}
\usepackage{braket}
\usepackage{verbatim}
\usepackage{float}
\usepackage{bbold}
\usepackage{color}
\usepackage{xcolor}
\usepackage{relsize}
\usepackage{amsthm}
\usepackage{enumerate}
\usepackage{soul,xcolor}
\usepackage[T1]{fontenc}
\usepackage[colorlinks=true ,urlcolor=blue,urlbordercolor={0 1 1}]{hyperref}

\newcommand{\beq}{\begin{eqnarray}}
\newcommand{\eeq}{\end{eqnarray}}

\newcommand{\bsp}{\begin{split}}
\newcommand{\esp}{\end{split}}

\newcommand{\be}{\begin{equation}}
\newcommand{\ee}{\end{equation}}

\begin{document}

\setstcolor{red}

\title{Fractional Fermi liquid in a generalized $t-J$ model}
\author{Ya-Hui Zhang}
\email[\href{mailto:zhangyh1992@gmail.com}{zhangyh1992@gmail.com}]{}
\affiliation{Department of Physics, Harvard University, Cambridge, MA, USA}
\author{Zheng Zhu}
\email[\href{mailto:zhuzheng@ucas.ac.cn}{zhuzheng@ucas.ac.cn}]{}
\affiliation{Kavli Institute for Theoretical Sciences, University of Chinese Academy of Sciences, Beijing 100190, China}
\affiliation{Department of Physics, Harvard University, Cambridge, MA, USA}

\date{\today}

\begin{abstract}
Inspired by the recent discovery of superconductivity in the nickelate  Nd$_{1-x}$Sr$_x$NiO$_2$, we study a generalized $t-J$ model  to investigate  the correlated phases induced by doping spin-one  Ni$^{2+}$ into a  spin $1/2$ Mott insulator formed by Ni$^{1+}$. Based on a three-fermion parton mean field analysis, we identify a robust fractional Fermi liquid (FL*) phase for almost every doping level. The FL* state is characterized by a small  Fermi pocket on top of  a spin liquid, which violates the Luttinger theorem of a conventional Fermi liquid and is an example of  a symmetric pseudogap metal. Furthermore, we verify our theory in one dimension through density matrix renormalization group (DMRG) simulations on both the generalized $t-J$ model and a two-orbital Hubbard model. The fractional Fermi liquid reduces to fractional Luttinger liquid (LL*) in one dimension, which is connected to the conventional  Luttinger liquid through a continuous quantum phase transition by tuning interaction strength. Our findings offer new insights into correlated electron phenomena in nickelate superconductors and other multi-orbital transition metal oxide with spin-triplet $d^8$ state.
\end{abstract}

\maketitle{}

 \section{Introduction}

 The search for the exotic states of matter beyond the Landau paradigm in realistic models and materials is one of the most important issues in the condensed matter field. In contrast to the conventional Fermi liquids, these exotic states host a set of intriguing properties such as the fractionalization, which is believed to be relevant to the interplay between charge, spin and orbital degrees of freedom but still to be fully understood. One prominent example is the  fractional Fermi liquid (FL*) emerged from doping a Mott insulator \cite{senthil2003fractionalized,senthil2004weak,paramekanti2004extending,mei2012luttinger,chowdhury2014density,punk2015quantum,feldmeier2018exact,zhang2020pseudogap}.
The FL* represents a set of fractionalized states which harbor a Fermi pockets on top of a ``spin liquid" background, as first proposed for the pseudogap phase in the hole doped cuprates\cite{lee2006doping,proust2019remarkable}.  Recent experiments of cuprates find evidence of a pseudogap metal with Hall number equal to $x$ instead of $1+x$ for doping level $x$ in the region $x<x^*\approx 0.19$\cite{badoux2016change}. Interestingly, no translation symmetry breaking order is found just below $x^*$\cite{proust2019remarkable}, suggesting the violation of the Luttinger theorem\cite{oshikawa2000topological}. To have a translation invariant pseudogap metal with small carrier density, fractionalization is necessary. Then the FL* becomes a promising candidate\cite{norman2010trend}. However, to our best knowledge, the FL* phase has not been identified as a ground state in any correspond models including Hubbard model and $t-J$ model\cite{leblanc2015solutions,jiang2019superconductivity,jiang2017holon}.

Theoretically, the FL* could be expected in the weak orbital coupling limit\cite{senthil2003fractionalized,senthil2004weak} of a two-orbital system. For instance, the orbital-selective Mott transition\cite{anisimov2002orbital,vojta2010orbital,pepin2007kondo} would give rise to a Mott localized orbital with local spin moment coupled to an orbital with itinerant Fermi pocket through a Kondo coupling $J_K$\cite{vojta2010orbital}. When $J_K \rightarrow 0$, the Fermi pocket decouples from the local spins, leading to a magnetic ordered metal or a FL* phase\cite{senthil2003fractionalized,senthil2004weak,tsvelik2016fractionalized,seifert2018fractionalized,hofmann2019fractionalized,danu2020kondo}, depending on whether the localized spin moment orders or not. Nevertheless, in the strong orbital coupling limit (i.e., $J_K \rightarrow +\infty$),  it is still unclear whether a FL* phase exists or not.  Previous search for such phase mainly focus on the hole doped cuprates, in which the doped hole enters the oxygen $p$ orbital and strongly couples to the spin moment in the copper $d_{x^2-y^2}$ orbital\cite{zhang1988effective}, but due to the formation of  Zhang-Rice singlet\cite{zhang1988effective}, the two orbitals become indistinguishable. Therefore, the existence of a FL* as a ground state of a lattice model and the corresponding materials beyond the orbital decoupling limit is still an open problem.  In addition, the nature of the quantum phase transition between a FL with a large Fermi surface and a FL* with a small Fermi surface is also of fundamental importance \cite{si2001locally,coleman2001fermi,senthil2004weak,pivovarov2004transitions,paul2008multiscale,gegenwart2008quantum,zhang2020deconfined}.

Experimentally, the recent discovery of the superconductivity in nickelate  Nd$_{1-x}$Sr$_x$NiO$_2$ \cite{li2019superconductivity} has brought new excitements to explore the above intriguing issues, which are now under intensive investigations~\cite{Hepting:2020p381,li2020superconducting,zeng:2020phase,Botana:2020p011024,Lechermann:2020p081110,Jiang:2019p201106,karp:2020,Hu:2019p032046,adhikary2020orbital,
Mi:2020p207004,Hu:2019p032046,zhang2020type,Zhang:2020p013214,Lang:2020doped,karp:2020many,Krishna:2020arxiv}. Physically if we dope holes into the spin $1/2$ Mott insulator in $d^9$ state (with one hole occupying the $d_{x^2-y^2}$ orbital), the resulting $d^8$ site (with two holes) may be in a spin-singlet or a spin-triplet state depending on the competition between the energy splitting of the two $e_g$ orbitals and the Hund's coupling.  If the Hund's coupling $J_H$ wins, the doped holes enter the $d_{z^2}$ orbital and couple with the spin $1/2$ moment from $d_{x^2-y^2}$ orbital through a large ferromagnetic  Kondo coupling  $J_K=-J_H$. This physics can be captured by a  generalized $t-J$ model on the square lattice as  first proposed by one of us\cite{zhang2020type}, which describes a set of transition metal oxides with two partially filled $e_g$ orbitals. Then the examination of this model would be a reasonable and realistic starting point to study the correlated phenomena in nickelates and other multi-orbital materials

Motivated by the above, we offer theoretical and numerical study of this generalized $t-J$ model, which is an effective model Hamiltonian of a two-orbital Hubbard model in the $J_H \rightarrow \infty$ limit and projected to a restricted Hilbert space with three spin-triplet doublon states and two spin $1/2$ singly occupied states\cite{zhang2020type}. By performing a self-consistent mean field calculation based on a three-fermion parton theory with $U(2)$ gauge structure, we find a  FL* ground state for every doping for a class of models with $SU(N)$ spin rotation symmetry. The essential point is that there can be two emergent spin $1/2$ fermions fractionalized from the spin-one doublon state.  Then one fermion become electron like and forms a small Fermi surface with volume $V_{FS}=x$, while the other fermion stays as a neutral spinon.  These two emergent fermions are different from the two original microscopic orbitals and they are now only weakly interacting with each other.   The emergence of two different effective orbitals at low energy is the key to have a stable FL phase in the strong coupling limit where there is no well-defined notion of microscopic orbital.  We test our theory in one dimension through DMRG simulation of the $t-J$ model.  We find clear numerical evidence of a 1D version of FL* phase, which we dub as fractional Luttinger liquid (LL*).   This result is striking given that only conventional Luttinger liquid has been found in the conventional $t-J$ model\cite{ogata1991phase,moreno2011ground}. We further find a direct quantum phase transition from large Fermi surface to small Fermi surface by tuning Hubbard $U$ and Hund's interaction $J_H$ together in  a two-orbital Hubbard model.  The charge compressibility diverges at the critical point.  We expect  more interesting physics and critical behaviors in higher dimension, which can be accessed in real experiments.

\section{Generalized $t-J$ model.}

In this section we introduce the generalized $t-J$ model.  We will first review the derivation of it for the spin $1/2$ case starting from a two-orbital Hubbard model, as done in Ref.~\onlinecite{zhang2020type}.  Then we will propose the $t-J$ model with a general $SU(N)$ spin.  The $N>2$ case is useful for performing mean field calculation in the next section.

\subsection{Derivation of the $t-J$ model for  Spin $1/2$}

 {We consider a transition metal oxide with active $3d$ orbitals. At the starting point, we assume that we have a good cubic symmetry and there is a large splitting between the two degenerate $e_g$ orbitals and the other three degenerate $t_{2g}$ orbitals.  We use the hole picture and add holes starting from the $d^{10}$ state. When the hole density $n<4$, we can just focus on the two $e_g$ orbitals. }   A general lattice Hamiltonian is

\begin{align}
	H&=H_K+\frac{U_1}{2}\sum_i n_{1;i}(n_{1;i}-1)+\frac{U_2}{2}\sum_i n_{2;i}(n_{2;i}-1)\notag\\
	&+U'\sum_i n_{1;i}n_{2;i}-2J_H \sum_i (\mathbf{S}_{1;i}\cdot \mathbf{S}_{2;i}+\frac{1}{4}n_{i;1}n_{i;2})
\label{eq:spin_orbital_model}
\end{align}
where $n_{a;i}$ is the density of the orbital $a$ at the site $i$. $a=1,2$ denotes the $d_{x^2-y^2}$ and the $d_{z^z}$ orbital respectively. $U_1$, $U_2$ are intra-orbital Hubbard interaction. $U'$ is the inter-orbital interaction. $J_H$ is the inter-orbital Hund's coupling.  It is expected that $U_1=U_2=U$ and $U-U'=2J_H$ 
{ in the limit with good cubic symmetry. Next we introduce a small crystal field which further splits the degeneracy of the two $e_g$ orbitals.  }

The kinetic energy is
\begin{align}
	H_K&=\sum_i \epsilon_{dd} n_{2;i}+ V \sum_i (d^\dagger_{i;1}d_{i;2}+h.c.) \notag\\
	&+ \sum_{\langle ij \rangle} t_{1;ij} d^\dagger_{1;i}d_{1;j}
	+ \sum_{\langle ij \rangle} t_{2;ij} d^\dagger_{2;i}d_{2;j}\notag\\
	&+\sum_{\langle ij \rangle} t_{12;ij}d^\dagger_{1;i}d_{2;j}+h.c. \notag\\
	&-\mu \sum_i  (d^\dagger_{i;1}d_{i;1}+d^\dagger_{i;2}d_{i;2})
	\label{eq:full_Hamiltonian}
\end{align}
where $\epsilon_{dd}$ is the splitting between the two $e_g$ orbitals.

The total density of particle $n=n_1+n_2$ is controlled by the chemical potential $\mu$ in the grand canonical ensemble.   For simplicity, we can also use the canonical ensemble and consider a fixed density $n=1+x$.  First let us focus on the doping at integer filling $n=1$. We will focus on the regime that $U,U',J_H>>t$.  At this limit, the system is in a strong Mott insulator with frozen density $n_i=1$ at each site.  We further assume the orbital energy splitting $\epsilon_{dd}>t$, which is usually true in quasi 1D and quasi 2D system without the cubic symmetry.  { However we still require that $\epsilon_{dd}<U,U',J_H$ and hence the splitting does not significantly influence the on-site interaction, which sets the largest energy scale.} Then the low energy physics of  Mott insulator  is essentially a spin $1/2$ model formed by the orbital $d_1$: $\ket{\uparrow}=d^\dagger_{i;1\uparrow}\ket{0}$ and $\ket{\downarrow}=d^\dagger_{i;1\downarrow} \ket{0}$.     The excitation $d^\dagger_{i;2} d_{i;1}$ is suppressed by the orbital splitting $\epsilon_{dd}$.

For simplicity, we label the state with $n_i=1$ as singlon and the state with $n_i=2$ as doublon.   The doublon state always have higher energy than the singlon state because of the large repulsion. The energy difference between the doublon and the singlon determines the Mott gap.  The exact energy of the doublon depends on its spin-orbital nature.   There are in total different $6$ different doublon states at each site $i$:  $\ket{1}= d^\dagger_{i;1\uparrow} d^\dagger_{i;1\downarrow} \ket{0}$, $\ket{2}= \frac{1}{\sqrt{2}}(d^\dagger_{i;1\uparrow}d^\dagger_{i;2\downarrow}+d^\dagger_{i;1\downarrow}d^\dagger_{i;2\uparrow})\ket{0}$, $\ket{3}=d^\dagger_{i;1\uparrow}d^\dagger_{i;2\uparrow}\ket{0}$, $\ket{4}=d^\dagger_{i;1\downarrow}d^\dagger_{i;2\downarrow}$, $\ket{5}= \frac{1}{\sqrt{2}}(d^\dagger_{i;1\uparrow}d^\dagger_{i;2\downarrow}-d^\dagger_{i;1\downarrow}d^\dagger_{i;2\uparrow})\ket{0}$ and $\ket{6}=d^\dagger_{i;2\uparrow}d^\dagger_{i;2\downarrow}\ket{0}$. It is easy to calculate the energy of these six doublon states to be: $U_1, \epsilon_{dd}+U'-J_H, \epsilon_{dd}+U'-J_H, \epsilon_{dd}+U'-J_H, \epsilon_{dd}+U', 2 \epsilon_{dd}+U_2$, where  the energy  is defined compared to that of the singlon state.  Given that $U_1=U_2=U$ and $\epsilon_{dd}>0$ in the hole picture, the last two doublon states have higher energy and can be ignored.   The first doublon is a spin-singlet and the next three doublon states are the spin-triplet.   In the literature these two groups are usally called the low spin and the high spin states of the $d^8$ configuration with two holes per site.   The energy difference between the low spin and high spin state is $\Delta=U-U'+J_H-\epsilon_{dd}=3J_H-\epsilon_{dd}$ if we take $U-U'=2J_H$.   If $\Delta<0$, the spin-singlet doublon is favored and we should get the conventional $t-J$ model upon doping.   In this paper we consider the case that $\Delta>0$, or equivalently $3J_H>\epsilon_{dd}$.

We consider the limit that $U-U'+J_H>>\epsilon_{dd}$ and $U,U', J_H >>t$ and $\epsilon_{dd}>>t$.    In this limit the doublon state is favored to be the three spin-triplet states.  Other doublon states can be ignored.   We need to emphasize the energy of the spin-triplet doublon is $\epsilon_{dd}+U'-J_H$ compared to the singlon, which is still much larger than the hopping and this sets the Mott gap at the integer filling $n=1$.   When we dope the Mott insulator, we need to introduce a large chemical potential term $\mu$ to compensate this large Mott gap.   We will take the simpler approach to fix the density to be $n=1+x$.  In this canonical ensemble approach, the energy difference between the doublon and the singlon is not important at all, as the number of doublons is fixed to be $x N_s$ and the number of singlon is fixed to be $(1-x)N_s$, where $N_s$ is the total number of sites in the system.   One may wonder why we do not consider the state with $n_i=0$.  To create an empty site, we must increase the number of doublon by $1$, which costs a large energy.   Therefore the low energy state when the density is fixed to be $n=1+x$ can only have singlon and doublon states.   This is a feature also shared by the conventional $t-J$ model.

For non-zero doping $x>0$, we can build a $t-J$ model  by projecting to a restricted Hilbert space with $5=2+3$ states per site.  The Hilbert space of each site is generated by five states: $\ket{\uparrow}=d^\dagger_{i;1\uparrow}\ket{0}, \ket{\downarrow}=d^\dagger_{i;1\downarrow}\ket{0}$, $\ket{S_z=1}=d^\dagger_{i;1\uparrow}d^\dagger_{i;2\uparrow}\ket{0}$, $\ket{S_z=0}=\frac{1}{\sqrt{2}}(d^\dagger_{i;1\uparrow}d^\dagger_{i;2\downarrow}+d^\dagger_{i;1\downarrow}d^\dagger_{i;2\uparrow})\ket{0}$, $\ket{S_z=-1}=d^\dagger_{i;1\downarrow}d^\dagger_{i;2\downarrow}\ket{0}$.  The first two are the spin $1/2$ singlon states and the last three are the spin-triplet doublon states.  We can project the physical operator into this restricted Hilbert space.  First, we find $P d_{i;1} P=0$ within this projected Hilbert space and hence $d_1$ does not enter the $t-J$ model. Here the operator $P$ projects into the restricted Hilbert space.   For the electron operator, we only need to keep $d_{i;2}$, which we also relabel as $c_i=d_{i;2}$.   Within the restricted Hilbert space, the electron operator is

\begin{align}
	c_{i;\uparrow}&= -\prod_{j<i}(-1)^{n_j}  (\ket{\uparrow}_i\bra{Sz=1}_i+\frac{1}{\sqrt{2}} \ket{\downarrow}_i \bra{S_z=0}) \notag\\
	c_{i;\downarrow}&= -\prod_{j<i}(-1)^{n_j}  (\ket{\downarrow}_i\bra{Sz=-1}_i+\frac{1}{\sqrt{2}} \ket{\uparrow}_i \bra{S_z=0})
\end{align}
where $c_{i}=d_{i;2}$.  $\prod_{j<i} (-1)^{n_j}$ is the Jordan-Wigner string to implement the fermion statistics.  For $2D$ system one can use any ordering of the lattice sites.  $n_i=c^\dagger_i c_i$ is the number of the doublon states:

\begin{equation}
	n_i=\sum_{m=-1,0,1} \ket{S_z=m}_i\bra{S_z=m}_i
\end{equation}

In our new $t-J$ model, both the doublon and the singlon carry spin.  We can define spin operator $\vec{s}_i$ for singlon and the spin operator $\vec{S}_i$ for the doublon.  They can be explicitly written in the basis we choose:

\begin{align}
\vec{s}_i&= \frac{1}{2} \sum_{\sigma \sigma'}\vec{\sigma}_{\sigma \sigma'}\ket{\sigma}_i\bra{\sigma'}_i \notag\\
\vec{S}_i&= \sum_{\alpha,\beta=1,0,-1} \vec{T}_{\alpha \beta} \ket{\alpha}_i \bra{\beta}_i
\end{align}
where $\vec{\sigma}$ is the usual Pauli matrix for spin $1/2$.  $\vec T$ is the $S=1$ spin operator within the three spin-triplet subspaces.   One can see that $\vec{s}_i$ and $\vec{S}_i$ commute with each other because they act on different subspaces.  The explicit value of the $\vec{T}$ matrix is listed below:

\begin{align}
T_z&=\begin{pmatrix} 1 & 0 & 0 \\ 0 &0 &0 \\ 0 & 0 &-1 \end{pmatrix} \notag\\
T_x&=\frac{1}{\sqrt{2}}\begin{pmatrix} 0 & 1 & 0 \\ 1 &0 &1 \\ 0 & 1 &0 \end{pmatrix} \notag\\
T_y&=\frac{1}{\sqrt{2}}\begin{pmatrix} 0 & -i & 0 \\ i &0 &-i \\ 0 & i &0 \end{pmatrix} \notag\\
\end{align}

After defining the operators, we can write down the $t-J$ model as:

\begin{align}
H_{t-J}&=H_t+H_J\\
H_t &=  - t \sum_{\langle ij \rangle}  c^\dagger_{i\sigma} c_{j\sigma}  +{\rm h.c.}\\
H_J &= \sum_{\langle ij \rangle}\big( J  { \vec s_i \cdot \vec s_j} + J_d {\vec S_i \cdot \vec S_j} + \frac{J'}{2}({ \vec s_i\cdot \vec S_j +\vec{S}_i\cdot \vec{s}_j})  \big)
\label{eq:t_J_model_main}
\end{align}

We needs to emphasize that all of the above terms are limited to the restricted Hilbert space. As a result, $c_i^\dagger c_j$ is not the usual hopping term. Instead, it is really an exchange term between the singlon and doublon states between $(i,j)$.    $J,J',J_d$ are super-exchange term at order of $t^2/U$. Their values depend on $\epsilon_{dd}, J_H, U, U'$, please see Ref.~\onlinecite{zhang2020type} for the derivation.  In this paper we take $J=J_d=J'$ just for simplicity.

Compared to the two-orbital Hubbard model, we can see that only one hopping term corresponding to $t=t_2$ is kept in the final $t-J$ model. This is because the orbital $d_{i;1}$ is frozen and we only have $c_i=d_{i;2}$ alive.  This also means that other values of hopping like $t_1,t_{12}, V$ in the two-orbital model does not matter when we take the $U,U',J_H>>t$, $J_H>>\epsilon_{dd}$ and $\epsilon_{dd}>>t$ limit.

\subsection{General $SU(N)$ case}

 We can generalize the above $t-J$ model with a general $SU(N)$ symmetry. We consider a Hubbard model with  two orbitals $d_{1;\alpha},d_{2;\alpha}$, where $\alpha=1,2,...N$.  We imagine that $d_2$ has larger energy than $d_1$.  The $n=1$ Mott insulator is formed by one $d_1$ electron per site, with a $SU(N)$ magnetic moment.    Then we dope the system to create doubly occupied site (doublon) with $n=2$.  We introduce a large Hund's coupling between $d_1,d_2$, so that the doublon state consists of one $d_1$ electron and one $d_2$ electron, forming a symmetric representation of $SU(N)$ (two row, one column in Young tableau).  For $N=2$, this is just a spin triplet.  We need to emphasize that the doublon state always have higher energy than the doublon state because of the Hubbard U, similar to the discussion in the previous subsection. A large $J_H$ is introduced to split the different doublon states, but it does not close the Mott gap.  The exact value of the doublon energy does not matter because we will fix the density of doublons in the canonical ensemble.

At each site, there are $N$ number of singly occupied (singlon) states  and $\frac{N(N+1)}{2}$ number of doublon states.  Thus the dimension of the Hilbert space at each site is $N+\frac{N(N+1)}{2}=\frac{N(N+3)}{2}$.   The $n=1$ state can be labeled as $\ket{\alpha}_i=d^\dagger_{i;1\alpha} \ket{0}$ with $\alpha=1,2,...,N$. Similarly the doublon state is labeled as $\ket{\alpha \beta}_i=\ket{\beta \alpha}_i=\frac{1}{2} F_{\alpha \beta}  (d^\dagger_{i;1\alpha} d^\dagger_{i;2\beta}-d^\dagger_{i;2 \alpha}d^\dagger_{1;\beta}\ket{0}$. $F_{\alpha\beta}=1$ when $\alpha=\beta$ and $F_{\alpha \beta}=\sqrt{2}$ when $\alpha \neq \beta$ are introduced as normalization factor.

Next we need to project the physical operators into this restricted Hilbert space. After projection, $P d_{i;1\alpha} P=0$ and $d_{i;2\alpha}$ becomes
\begin{align}
c_{i;\alpha}=- \prod_{j<i}(-1)^{n_j}  \sum_{\beta}  G_{\alpha \beta}\ket{\beta}_i \bra{\alpha \beta}_i
\end{align}
where $G_{\alpha \beta}=1$ when $\alpha=\beta$ and $G_{\alpha \beta}=\frac{1}{\sqrt{2}}$ when $\alpha \neq \beta$.  Here we define $c_{i;\alpha}=d_{i;2\alpha}$.

We can also define the spin operator for the singlon and doublon sites.  For singlon, the spin operator is
\begin{equation}
	s^\alpha_\beta(i)= \ket{\alpha}_i \bra{\beta}_i
\end{equation}
For doublon, the spin operator is
\begin{equation}
	S^\alpha_\beta(i)= P  \left(d^\dagger_{i;1\alpha}d_{i;1\beta}+d^\dagger_{i;2\alpha}d_{i;;2\beta} \right) P
\end{equation}
where $P$ is the projection operator to the doublon state.  One can write down the terms after projection.  Fortunately this is not necessary for our purpose.

With the above definition of Hilbert space and physical operators, the generalized $t-J$ model can be written as
\begin{align}
H_{t-J}&=H_t+H_J\\
H_t &=  - \sum_{\langle ij \rangle}t_{ij}c^\dagger_{i\alpha} c_{j\alpha} +{\rm h.c.}\\
H_J &= \sum_{\langle ij \rangle} \frac{J}{2}   s^\alpha_\beta(i) s^\beta_\alpha(j) + \frac{J_d}{2} S^\alpha_\beta(i) S^\beta_\alpha(j)\notag\\
&~~~ + \frac{J'}{4}\big( s^\alpha_\beta(i) S^\beta_\alpha(j)+S^\alpha_\beta(i) s^\beta_\alpha(j) \big)
\label{eq:t_J_model_main}
\end{align}
The normalization factor in front of the spin-spin coupling is chosen so that the term reduces to the traditional $\vec S \cdot \vec S$ form for $N=2$, as written in Ref.~\onlinecite{zhang2020type}. For simplicity we will consider $J'=J_d=J$ in this paper.

\section{Three-fermion parton theory.}

To deal with a restricted Hilbert space, it is easier to work with parton theory.  In the conventional spin $1/2$ $t-J$ model, one can create the singlon state with an Abrikosov fermion operator $f^\dagger_\sigma$ and create the spinless doublon with a slave boson operator $b^\dagger$\cite{lee2006doping}.  In our case, both the singlon and the doublon carry spin. The singlon state can still be generated by a fermion operator: $\ket{\alpha}_i=f^\dagger_{i;\alpha}\ket{0}$. The doublon is in a representation with a huge dimension $d=\frac{N(N+1)}{2}$.  This symmetric representation can be generated by two-orbital fermions with a $U(2)$ gauge constraint\cite{zhang2020type}: $\ket{\alpha \beta}_i= \frac{1}{2} F_{\alpha \beta}\epsilon_{ab} \psi^\dagger_{i;a\alpha}\psi^\dagger_{i;b\beta} \ket{0}$.

Then the electron operator is
\begin{equation}
	c_{i;\alpha}= \frac{1}{2} \epsilon_{ab} f^\dagger_{i;\beta}   \psi_{i;a\alpha}\psi_{i;b\beta}
\end{equation}
The singlon spin operator and doublon spin operator can also be written as
\begin{align}
s^\alpha_\beta(i)&= f^\dagger_{i;\alpha}f_{i;\beta}\notag\\
S^\alpha_\beta(i)&= \sum_{a=1,2}\psi^\dagger_{i;a\alpha}\psi_{i;a \beta}
\end{align}
For convenience we define a spinor $\Psi_{i;\alpha}=(\psi_{i;1\alpha},\psi_{i;2\alpha})^T$ and label the Pauli matrices $\tau_a$ acting on this spinor. The above operators becomes the correct physical operator when we implement the constraint:
\begin{align}
&f^\dagger_{i;\alpha}f_{i;\alpha}+\frac{1}{2} \Psi^\dagger_{i;\alpha}\Psi_{i;\alpha}=1\notag\\
& \Psi^\dagger_{i;\alpha} \vec \tau \Psi_{i;\alpha}=0
\end{align}
The above two constraints generate a $(U(1) \times SU(2))/Z_2=U(2)$ gauge symmetry. $U(1)$ fixes the first constraint above while $SU(2)$ rotates in the $(\psi_1, \psi_2)$ space.  With the parton theory, we can rewrite the original Hamiltonian as shown  in the appendix.

\textbf{FL* phase.} We can write down mean field ansatz using the three-fermion parton theory by decoupling the original Hamiltonian to bilinear terms of the partons. For simplicity we focus on the translation-invariant ansatz.
\begin{align}
	H_M&=-t_f \sum_{\langle ij \rangle}(f^\dagger_{i;\alpha}f_{j;\alpha}+h.c.)-t_{\psi;ab} \sum_{\langle ij \rangle}(\psi^\dagger_{i;a\alpha}\psi_{j;b\alpha}+h.c.)\notag\\
	&-\mu_f \sum_i f^\dagger_{i;\alpha}f_{i;\alpha}-\mu_{ab} \sum_i \psi^\dagger_{i;a \alpha}\psi_{i;b\alpha}\notag\\
	&-\Phi_0 \sum_i (f^\dagger_{i;\alpha} \psi_{i;1\alpha}+h.c.) -\Phi_a \sum_{ij }(f^\dagger_{i;\alpha}\psi_{j;a\alpha}+h.c.)
\end{align}
In the above we did not include $f^\dagger_{i;\alpha} \psi_{i;2\alpha}$ because we can always use the local $SU(2)$ gauge transformation to remove it.  Here we fix the gauge so that only $\psi_1$ has an on-site coupling to $f$.  $\psi_2$ can only hybridize $f$ through nearest neighbor coupling $\Phi_2$.  Chemical potentials are introduced to fix $\langle n_{\psi_1} \rangle=\langle n_{\psi_2} \rangle=1-\langle n_f \rangle=x$ and $\langle \psi^\dagger_{i;1\alpha}\psi_{i;2\alpha} \rangle=0$.

\begin{figure}[ht]
\centering
\includegraphics[width=0.5\textwidth]{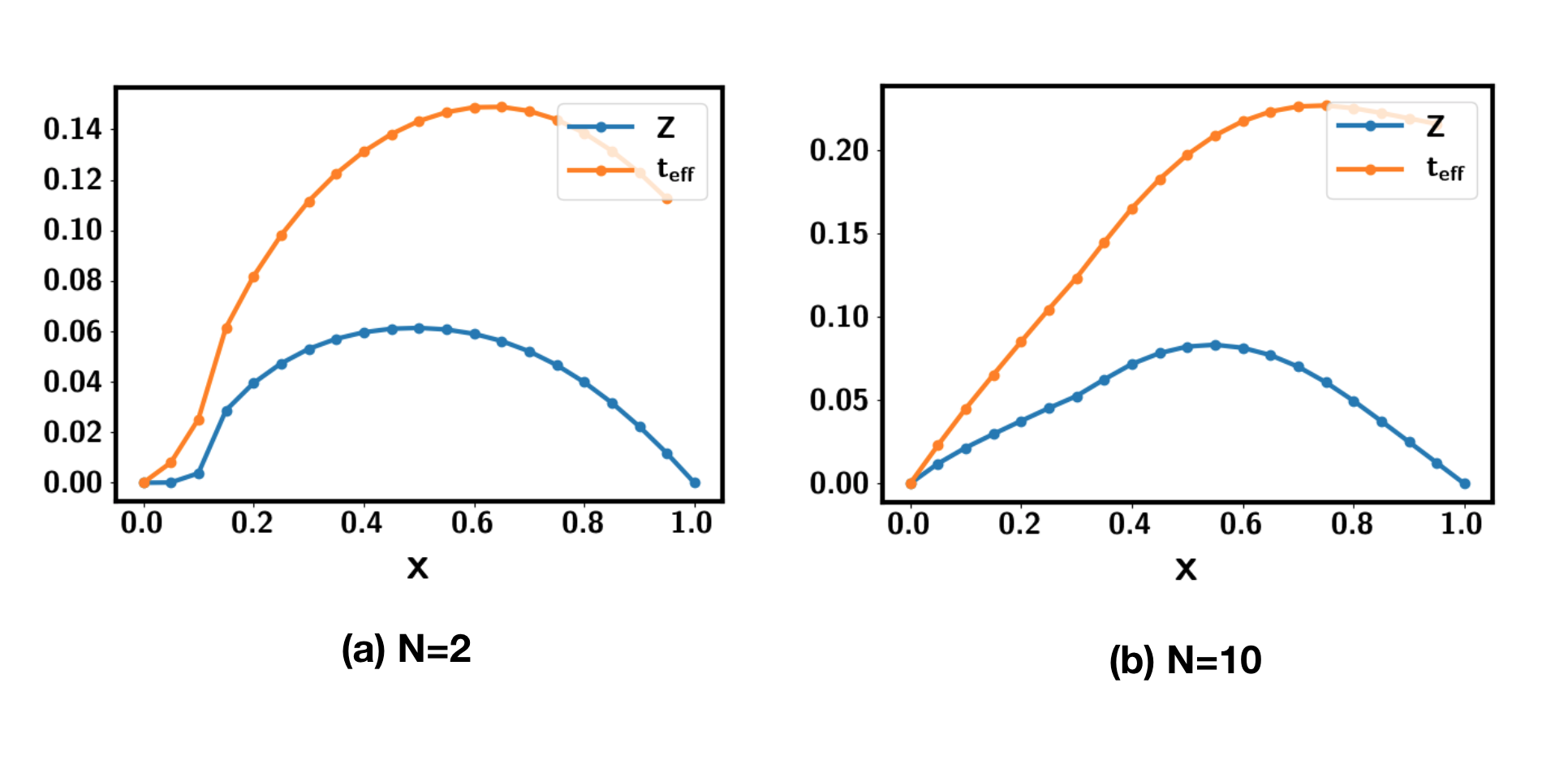}
\caption{Mean field ansatz at zero temperature with doping $x$ for $N=2$ and $N=10$ on square lattice.  We used the parameter $t=1$, $J=J_d=J'=0.5$.  $t_{eff}=t_{\psi;22}$ is the hopping of $\psi_2$ is in unit of $t$. $Z=|\frac{1}{2}\langle \sum_{\alpha}f^\dagger_{i;\alpha} \psi_{i;1\alpha} \rangle|^2$. The hoppings of  the spinons $f,\psi_1$ are listed in Fig.~\ref{fig:mean_field_append}.}
\label{fig:mean_field}
\end{figure}

We solve the self-consistent equations numerically on square lattice (please see details in the Appendix.~\ref{append:three_fermion}). A list of the obtained mean field parameters is shown in Fig.~\ref{fig:mean_field} and in Fig.~\ref{fig:mean_field_append}.  At zero temperature, we find that $\Phi_0 \neq 0, \Phi_1 \neq 0$ and $t_f\neq0, t_{\psi,11} \neq 0, t_{\psi,22} \neq 0$, but $\Phi_2=t_{\psi;12}=0$.  As shown in Ref.~\onlinecite{zhang2020type} and in the appendix, this ansatz describes a FL* phase.    We can see this  from the original definition $c_{i;\alpha}= \frac{1}{2} \epsilon_{ab} f^\dagger_{i;\beta}   \psi_{i;a\alpha}\psi_{i;b\beta} $.  Because $ \frac{1}{2}\sum_{\alpha}\langle f^\dagger_{i;\alpha} \psi_{i;1\alpha} \rangle=\sqrt{Z} \neq  0$, we can identify $c_{i;\alpha}=\sqrt{Z} \psi_{i;2\alpha}$.  This implies the Green function $G_c(\omega, \mathbf k)=Z G_{\psi_2}(\omega, \mathbf k)$ and $Z$ can be identified as quasiparticle residue.   In contrast, $f$ and $\psi_1$ do not have overlap with $c$. Actually $f,\psi_1$ still couples to a $U(1)$ gauge field and should be interpreted as neutral spinons (see the Appendix.~\ref{append:three_fermion}).

The final phase is a fractional Fermi liquid. There is a small Fermi pocket formed by $\psi_2$, whose volume is $V_{FS}=\frac{x}{N}$ for each flavor.  The violation of the Luttinger theorem $V^0_{FS}=\frac{1+x}{N}$ is compensated by the existence of a spin liquid formed by $f,\psi_1$. In our ansatz the spinons just form a spinon Fermi surface coupled to a $U(1)$ gauge field.

If we focus on translation invariant ansatz, the FL* phase is the only ansatz we find at any doping level $x$ for the parameter $t=1,J=0.5$.  It is quite remarkable that a Fermi liquid with large volume is not favored according to our mean field analysis.  We need to emphasize that our analysis here can not rule out more conventional phases such as antiferromagnetic metal, which does not fall in the translation invariant ansatz we restrict to.   As other parton theories such as slave boson theory, our parton mean field analysis is just an approximating method and it does not necessarily prove that the FL* phase is the ground state of the proposed $t-J$ model. However, it at least proves that FL* phase can in principle exist as a legitimate state.  The energetical problem of whether it is indeed favored for a given microscopic model can only be determined through numerical simulation or compared to experimental results.  In the following sections we will provide evidence that a FL* phase as described by the parton mean field theory in this section is indeed the ground state for the $t-J$ model in one dimension.  This of course does not prove the existence of it in 2D. But combination of mean field theory here and the 1D numerics at least suggests FL* phases is a strong candidate for the $t-J$ model proposed in this paper.

\section{Numerical evidence of FL* in 1D.}

\begin{figure}
\centering
\includegraphics[width=0.45 \textwidth]{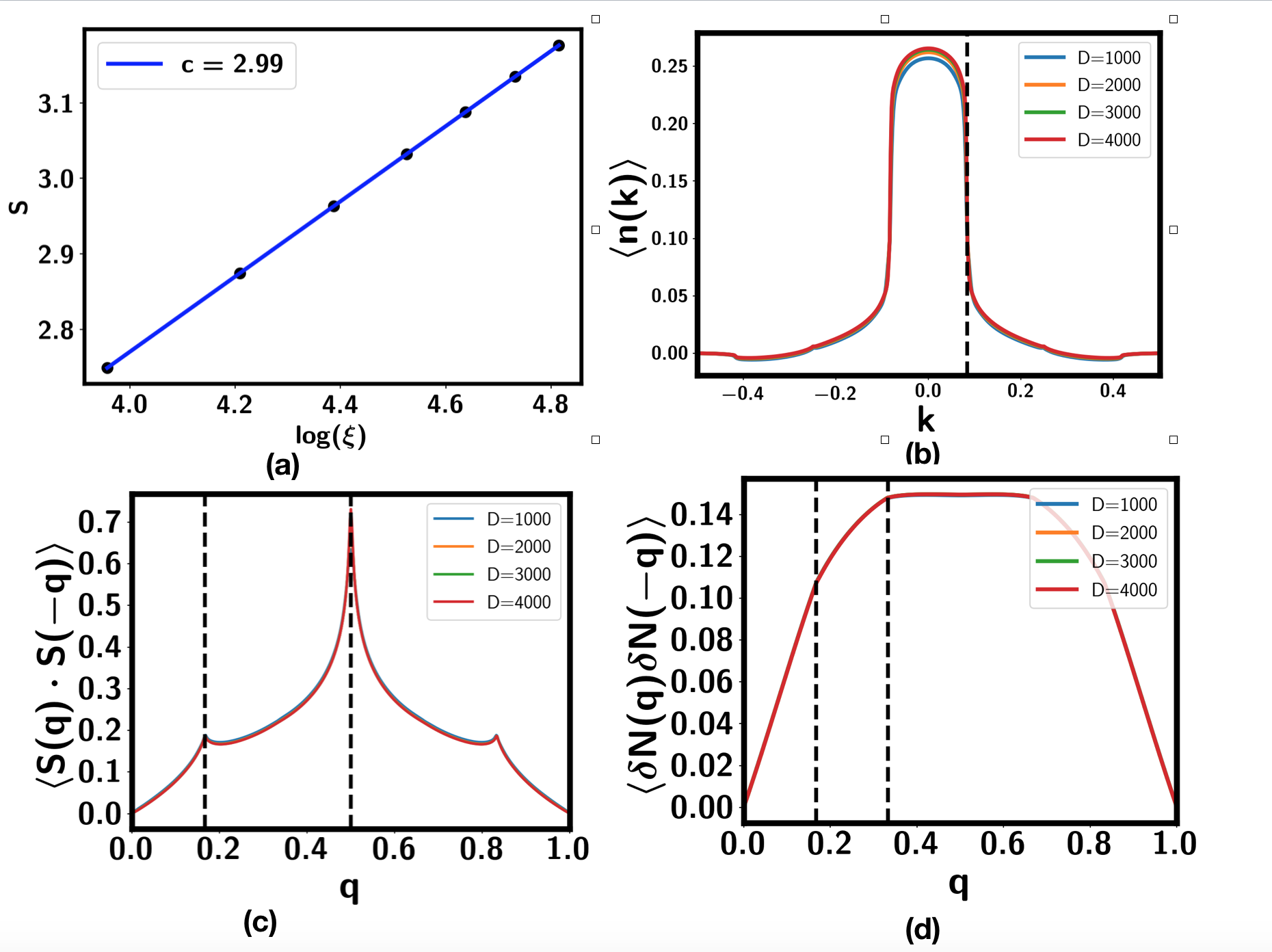}
\caption{Results for $x=\frac{1}{3}$ from iDMRG. We used the $t-J$ model as defined in Eq.~\ref{eq:t_J_model_main} with parameter $t=2J=1$ and $J_d=J'=J$.  The momentum is in unit of $2\pi$. (a) We vary the bond dimension $D$ from $1000$ to $4000$ to fit the central charge from $S=\frac{c}{6} \log \xi$, where $\xi$ is the correlation length. The obtained central charge is $c=2.99$. (b) Momentum distribution function $n(\mathbf k)=\langle c^\dagger(\mathbf k)c(\mathbf k)\rangle$. The dashed line is at $k_F^*=\frac{x}{4} 2\pi$. (c) Spin structure factor with peaks at $\mathbf q=2 k_F^*$ and $\mathbf q=\pi$. (d) Density density correlation function with weak discontinuities at $2k_F^*$ and $4 k_F^*$. }
\label{fig:idmrg}
\end{figure}

We simulate the $t-J$ model with $N=2$ in one dimension using DMRG.  Figures~\ref{fig:idmrg} show the results at filling $x=\frac{1}{3}$ from infinite DMRG (iDMRG). The momentum distribution $n(\mathbf k)$ clearly shows a small Fermi surface with size $2k_F^*=\frac{x}{2} 2\pi$ [see Fig.~\ref{fig:idmrg}(b)]. This small Fermi surface is further confirmed by the density-density correlation with discontinuities at $\mathbf q=2k_F^*$ in Fig.~\ref{fig:idmrg}(d).  We do not find any feature at $2k_F=\frac{1+x}{2} 2\pi$ corresponding to the large Fermi surface according to the Luttinger constraint. In Fig.~\ref{fig:idmrg}(c), the spin-spin correlation function shows two peaks at both $\mathbf q=2k_F^*$ and $\mathbf q=\pi$.  The first peak is apparently from the small Fermi surface. The mode at $\mathbf q=\pi$ is charge neutral because it does not show up in the density-density correlation function and the electron distribution.  Therefore we conclude that there is a small Fermi surface coexisting with another spin mode at $\mathbf q=\pi$. In total there are three modes, consistent with the result of the central charge $c\approx 3.0$ fit from  the entanglement entropy [see Fig.~\ref{fig:idmrg}(a)].

In addition to the three elementary modes, we can also find small features at other momentums, which correspond to bound states of the elementary excitations. This is a feature shared by the conventional Luttinger liquid.  For example, we can see feature for $n(\mathbf k)$ at $k=3 k_F^* =0.25 \times 2\pi$.  Such a $3k_F$ excitation is well known in the conventional Luttinger liquid.   In our case, because the spin excitation has another mode at $q=\frac{1}{2} 2\pi$ in addition to $q=2k_F^*$, an electron at momentum $-k_F^*$ can be scattered by this spin mode to $-k_F^*+\frac{1}{2} 2\pi \approx 0.416 \times  2\pi$ for $x=\frac{1}{3}$.  We indeed see a small feature at such momentum, as evidence for this higher order excitation.  In the Appendix.~\ref{append:dmrg_t_J} we show more detailed scaling with the bond dimension $D$ at $k=-k_F+\frac{1}{2} 2\pi$ and $k=k_F^*$. We find the rapid decrease of $n(\mathbf k)$ at $k=k_F^*$ becomes steeper when increasing $D$, while the feature at $\frac{1}{2} 2\pi -k_F^*$ actually becomes suppressed by larger $D$.  This suggests that the intensity of this higher order excitation at $\frac{1}{2}2\pi -k_F^*$ is much smaller than the elementary excitation at $k_F$.  In the results of finite DMRG at various doping levels, we do not find these other small features in addition to the dominant singularity at $k_F^*$ in $n(\mathbf k)$.  Therefore we believe  that the dominant excitation at $n(\mathbf k)$ is the one at $k_F^*$, consistent with our interpretation of a small Fermi surface with volume $2k_F^*$.   This is further supported by the fact that we do not find any feature at these other momentums in $\langle \mathbf{S}(\mathbf q)\cdot \mathbf{S}(-\mathbf q)\rangle$.  It is interesting to study the quantitative behavior of these higher order excitations, but it requires a more precise calculation as they are not robust. We leave a detailed analysis to future.

In the Mott insulator at $x=0$, there is a gapless spin mode at $\mathbf q=\pi$, which is described by the $SU(2)_1$ conformal field theory (CFT) and can be thought as a spin liquid with "spinon Fermi surface" in one dimension.  Our numerical results then suggest that the doped holes just form a small Fermi surface, which  coexists together with the "spin liquid" part in the Mott insulator.  This is exactly the behavior of a fractional Fermi liquid described in the parton theory.  Let us also comment on how this phase is compatible with the Lieb-Schultz-Mattis (LSM) constraint in one dimension\cite{lieb1961two,yamanaka1997nonperturbative}.  The LSM constraint says that for symmetric phase there must be a gapless model at crystal momentum $\mathbf Q=2\pi \nu$, where $\nu=\frac{1+x}{2}$ is the filling per spin\cite{yamanaka1997nonperturbative}.  In conventional Luttinger liquid, this gapless mode corresponds to $2k_F$ excitation.  However, it is possible that this required gapless mode is  fractionalized to two elementary modes. In our case, this  required gapless mode is formed by a bound state of a neutral mode with momentum $\pi$ from the spin liquid part and the $2k_F^*=\frac{x}{2} 2\pi$ of the small Fermi surface: $\mathbf Q=2\pi \nu= \pi+ 2k_F^*$.    One can see that the existence of a neutral spin liquid sector can shift the Luttinger constraint of the Fermi surface volume by $1/2$ Brillouin Zone (BZ). We dub such a phase fractional Luttinger liquid (LL*), as an analog of the FL* phase in higher dimensions.

\section{Small to large Fermi surface transition.}

\begin{figure}[htp]
\centering
\includegraphics[width=0.5\textwidth]{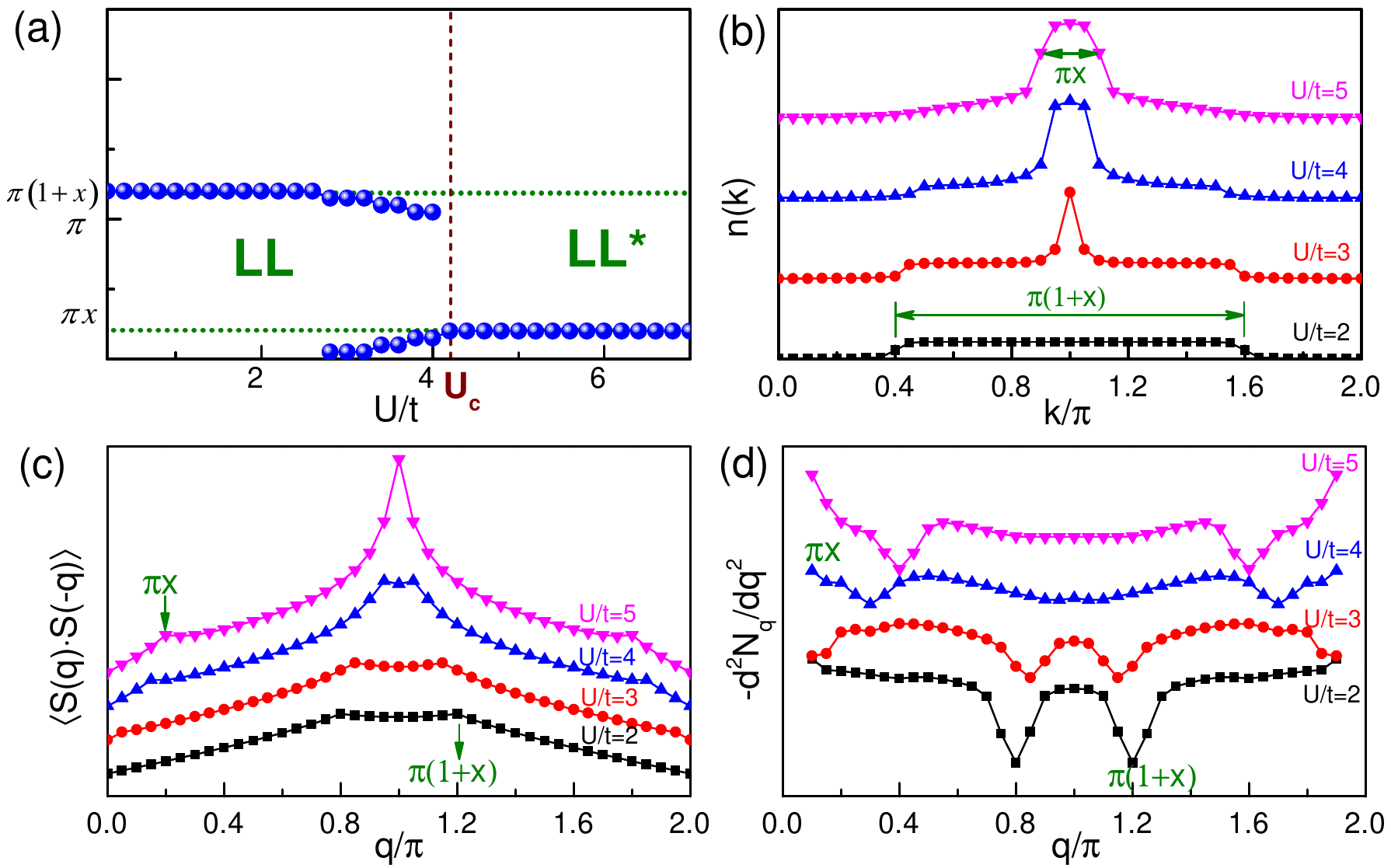}
\caption{Results of the two-orbital Hubbard model defined in Eq.~\ref{eq:spin_orbital_model} at $x=0.2$. We tune $U$ and Hund's coupling $J_H$ together with fixed ratio $U=4J_H, U'=2J_H$. For the hopping defined in Eq.~\ref{eq:full_Hamiltonian}, we use $t_1=t_2=V=1$, $\epsilon_{dd}=2$ and $t_{12}=0$.  In the electron picture, the filling is fixed to $\nu_T=3-x$, or in the hole picture $\nu_T=1+x$. Panel (a) shows the Fermi momentum in the conventional LL phase at smaller $U$ side and the fractional LL phase at large $U$ side. The Fermi momentum in these two phases are identified from the consistent evidence, including the sudden jump in momentum distribution in (b), the kinks in the spin structure factor in (c) and the second order derivative of the charge density structure factor in (d).
}
\label{fig:Hubbard}
\end{figure}

The generalized $t-J$ model can be derived in the $U,J_H>>t$ limit of a two-orbital Hubbard model (see the appendix).  In the weak coupling limit, the ground state must be a conventional Luttinger liquid (LL) phase with large Fermi surface.  Therefore we can study a LL to LL* transition tuned by $U$.

We simulate the two-orbital Hubbard model using finite DMRG at $x=0.2$ and show results in Fig.~\ref{fig:Hubbard}.  At $U=0$, there is a single Fermi surface with $2k_F=\frac{1+x}{2}$, formed mainly by $d_1$ orbital.  Then when $U>3$, the Fermi surface splits to two, presumably because the effective energy of the orbital $d_2$ is renormalized by the interaction and becomes smaller. But the total Fermi surface volume still satisfies the usual Luttinger constraint.  Then above a critical value $U_c$, one of the two Fermi surface becomes half-filled and gets a Mott gap, resulting in a LL* phase.

\begin{figure}[ht]
\centering
\includegraphics[width=0.4\textwidth]{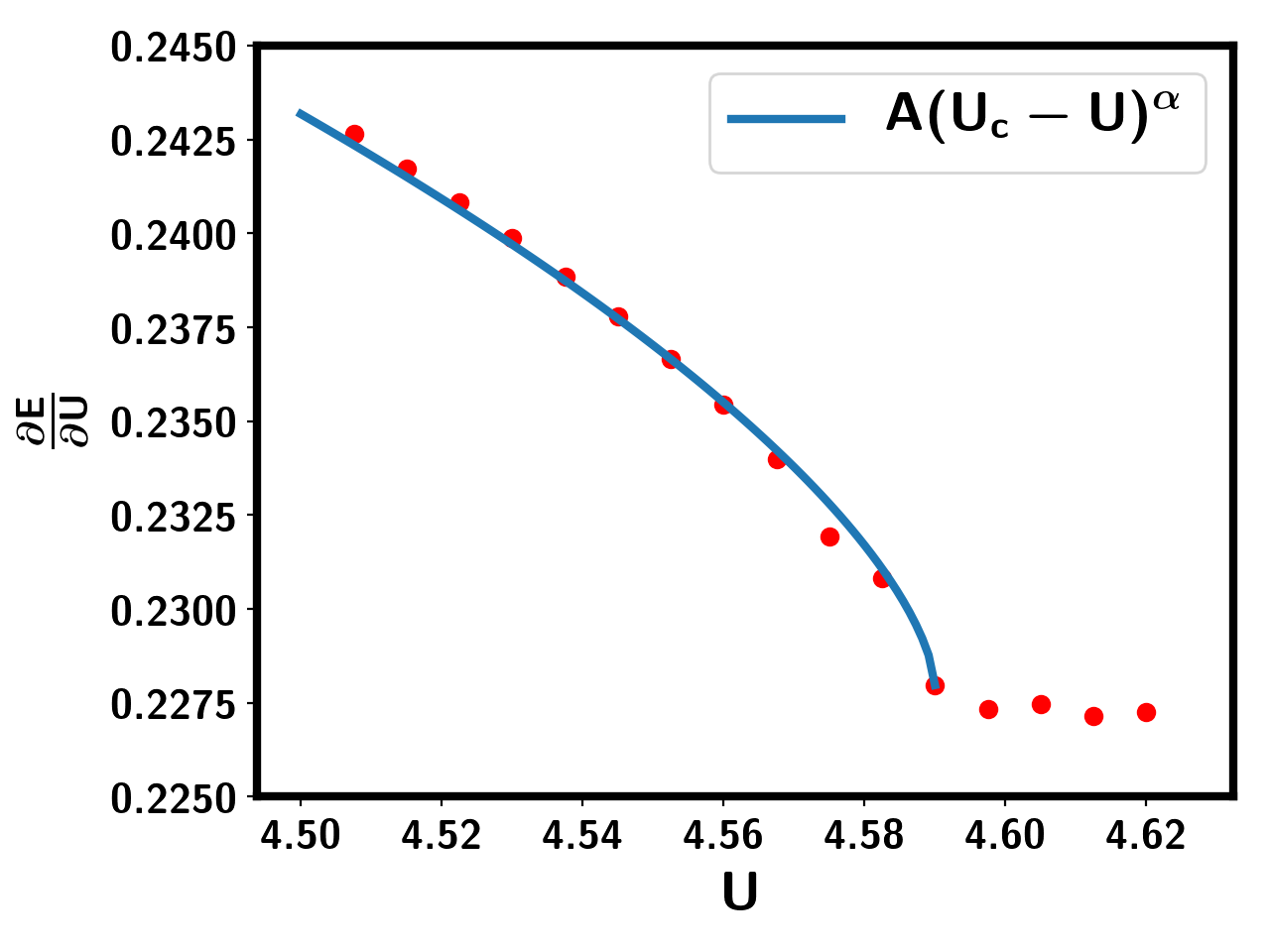}
\caption{$\frac{\partial E}{\partial U}$ from iDMRG with bond dimension $D=5000$ for doping $x=\frac{1}{3}$. The parameters is the same as in Fig.~\ref{fig:Hubbard}, we choose the doping $x=\frac{1}{3}$ because it is easier for iDMRG. $\frac{\partial E}{\partial U}$ is continuous, implying a continuous phase transition.  For $U<U_c$, $\frac{\partial E}{\partial U}$ can be fit with $A(U_c-U)^\alpha+C$ with $\alpha \approx 0.64$. }
\label{fig:singularity}
\end{figure}

If we ignore the small pocket, the critical pocket goes through a chemical potential tuned Mott transition with chemical potential $\mu-\mu_c \propto -(U-U_c)$.  It is known that for the chemical potential tuned transition in one-orbital Hubbard model\cite{giamarchi2003quantum}, $\langle n \rangle=-\frac{\partial E}{\partial \mu}=\begin{cases} A \sqrt{\mu-\mu_c}+1, & \mu>\mu_c, \\ 1 & \mu<\mu_c \end{cases}$.  Therefore, in the orbital-selective Mott transition picture, we expect $\frac{\partial E}{\partial_U}=\begin{cases}A (U_c-U)^\alpha+C, & U<U_c, \\ C & U>U_c \end{cases}$, where $\alpha=\frac{1}{2}$. At the critical point, there is also a divergence of the charge compressibility $\kappa \sim \frac{1}{(U_c-U)^{1-\alpha}}$.  Next we check this directly in numerical simulation.  As finite DMRG seems to suffer from a problem of discontinuous momentum jump due to finite size, we use iDMRG to examine the exponent around $U_c$. We indeed find a singularity for $\frac{\partial E}{\partial U}$ as shown in Fig.~\ref{fig:singularity}, but the fitted exponent is $\alpha \approx 0.64$, which is larger than that expected in the decoupling picture. Given the numerical noise in our calculation at the maximal bond dimension $D=5000$ limited by our computational resource, it is not clear whether the discrepancy is just from numerical error or actually implies a new universality class. In our model, the two orbitals $d_1,d_2$ are strongly coupled by $J_H$, and thus it may be possible that the coupling to the small pocket modifies the exponent of the critical Fermi surface.  We leave a systematic study to a future paper.

\section{Conclusion.}

In summary, we find a symmetric pseudogap metal with small Fermi surface in a generalized $t-J$ model based on parton theory and DMRG simulation in one dimension.  This generalized $t-J$ model can be realized in certain transition-metal-oxides, such as nickelates. It is known that Ni$^{2+}$ ion is in a spin-triplet state formed by the two $e_g$ orbitals in many cases, then doping spin $1/2$ Ni$^{1+}$ into a spin-one Mott insulator formed by Ni$^{2+}$ will realize our $t-J$ model. In one dimension, this can be achieved by doping the Haldane chain formed by $d^8$ state\cite{kojima1995musr}.  Recently, superconductivity was found in a quasi 2D nickelate Nd$_{1-x}$Sr$_x$NiO$_2$\cite{li2019superconductivity}.  It is still not clear whether the doped site is in a spin singlet or spin triplet state.  If spin-triplet is favored, then our model is relevant for Nd$_{1-x}$Sr$_x$NiO$_2$\cite{zhang2020type} and our theory suggests a non-trivial metallic phase with only small pocket above the superconductor $T_c$.  We need to emphasize that in $2D$ we can not rule out other more conventional state such as an antiferromagnetic metal.  A numerical calculation in $2D$ is needed to determine whether the FL* phase, or more conventional phase, is favored in the generalized $t-J$ model.  Numerical simulation in two dimension is much harder and we hope the current paper can motivate more numerical efforts on simulating this new $t-J$ model in various 2D lattices.

\textbf{Acknowledgement}
We thank Ashvin Vishwanath for discussions and previous collaboration. The DMRG simulations of the $t-J$ model and iDMRG simulation of the two-orbital Hubbard model were performed using the TeNPy Library (version 0.4.0)\cite{hauschild2018efficient}.  YHZ was supported by the Simons Collaboration on Ultra-Quantum Matter, which is a grant from the Simons Foundation (651440, AV, SS) and NSF Grant  DMR-2002850. ZZ was supported by the National Natural Science Foundation of China (Grant  No. 12074375), the Fundamental Research Funds for the Central Universities, the start-up funding of KITS at UCAS, and the Strategic Priority Research Program of CAS (No. XDB33000000).

\appendix

\onecolumngrid

\section{Three-fermion parton theory of the generalized $t-J$ model \label{append:three_fermion}}

 In our three-fermion parton theory, the singlon state is generated by a fermion operator: $\ket{\alpha}_i=f^\dagger_{i;\alpha}\ket{0}$. The doublon is represented as: $\ket{\alpha \beta}_i= \frac{1}{2} F_{\alpha \beta}\epsilon_{ab} \psi^\dagger_{i;a\alpha}\psi^\dagger_{i;b\beta} \ket{0}$.

To recover the physical Hilbert space, we need to implement the constraint:

\begin{align}
&f^\dagger_{i;\alpha}f_{i;\alpha}+\frac{1}{2} \Psi^\dagger_{i;\alpha}\Psi_{i;\alpha}=1\notag\\
& \Psi^\dagger_{i;\alpha} \vec \tau \Psi_{i;\alpha}=0
\end{align}
where $\Psi_{i;\alpha}=(\psi_{i;1\alpha},\psi_{i;2\alpha})^T$.

\subsection{Gauge symmetry and the Higgs phases}
The above two constraints generate a $(U(1) \times SU(2))/Z_2=U(2)$ gauge symmetry. $U(1)$ fixes the first constraint above while $SU(2)$ rotates in the $(\psi_1, \psi_2)$ space.  The $U(1)$ gauge symmetry acts as: $f_{i;\alpha}\rightarrow e^{i 2\alpha_c(i) }f_{i;\alpha}, \Psi_{i;\alpha}\rightarrow e^{i \alpha_c(i)}\Psi_{i;\alpha}$. The $SU(2)$ acts as $f_{i;\alpha}\rightarrow f_{i;\alpha}, \Psi_{i;\alpha}\rightarrow U_i \Psi_{i;\alpha}$, where $U_i \in SU(2)$.  $U(1)$ and $SU(2)$ share a $Z_2$ center: $f_{i;\alpha}\rightarrow f_{i;\alpha}, \Psi_{i;\alpha}\rightarrow -\Psi_{i;\alpha}$, so the final gauge structure is $(U(1)\times SU(2))/Z_2=U(2)$.  The $U(2)$ has an Abelian subgroup $U(1)\times U(1)$, which acts as $\psi_{i;1\alpha}\rightarrow  \psi_{i;1\alpha}e^{i\alpha_{i;1}}, \psi_{i;2\alpha} \rightarrow \psi_{i;2\alpha}e^{i\alpha_{i;2}}, f_{i;\alpha} \rightarrow f_{i;\alpha} e^{i(\alpha_{i;1}+\alpha_{i;2})}$.   Basically if we label the corresponding two $U(1)$ gauge fields  as $a_1$ and $a_2$, then $\psi_1$ couples to $a_1$, $\psi_2$ couples to $a_2$ and $f$ couples to $a_1+a_2$.

Let us also discuss the coupling to the physical gauge field $A$.  We can assign charge in the following way: $\psi_1$ and $\psi_2$ carries $\frac{1}{2}$ charge and $f$ is neutral.    So finally $f$ couples to $a_1+a_2$, $\psi_1$ couples to $a_1+\frac{1}{2}A$ and $\psi_2$ couples to $a_2+\frac{1}{2}A$.  This charge assignment can be shifted if we redefine $a_1$ and $a_2$.  Therefore the physical charge of the partons is not well-defined unless the internal $U(1)$ gauge field is higgsed.

\subsection{Self consistent mean field calculation}

The original Hamiltonian of the generalized $t-J$ model with $SU(N)$ spin can be rewritten using the three-fermion partons:

\begin{align}
H&= \frac{1}{4}t \sum_{\langle ij \rangle} \epsilon_{ab}\epsilon_{a'b'} \psi^\dagger_{i;b\beta}\psi^\dagger_{i;a\alpha}\psi_{j;a'\alpha}\psi_{j;b'\beta'} f^\dagger_{j;\beta'}f_{i;\beta}+h.c. \notag\\
&~~-\frac{1}{2} J \sum_{\langle ij \rangle} f^\dagger_{i;\alpha}f_{j;\alpha}f^\dagger_{j;\beta}f_{i;\beta} \notag\\
&~~-\frac{1}{2} J_d \sum_{\langle ij \rangle} \psi^\dagger_{i;a\alpha}\psi_{j;b\alpha} \psi^\dagger_{j;b\beta}\psi_{i;a\beta}\notag\\
&~~-\frac{1}{4}J' \sum_{\langle ij \rangle} (f^\dagger_{i;\alpha}\psi_{j;a\alpha} \psi^\dagger_{j;a\beta} f_{i;\beta}+\psi^\dagger_{i;a\alpha}f_{j;\alpha} f^\dagger_{j;\beta}\psi_{i;a \beta})
\end{align}

We can then obtain a mean field ansatz through decoupling:

\begin{align}
	H_M&=-t_f \sum_{\langle ij \rangle}(f^\dagger_{i;\alpha}f_{j;\alpha}+h.c.)-t_{\psi;ab} \sum_{\langle ij \rangle}(\psi^\dagger_{i;a\alpha}\psi_{j;b\alpha}+h.c.)\notag\\
	&-\mu_f \sum_i f^\dagger_{i;\alpha}f_{i;\alpha}-\mu_{ab} \sum_i \psi^\dagger_{i;a \alpha}\psi_{i;b\alpha}\notag\\
	&-\Phi_0 \sum_i (f^\dagger_{i;\alpha} \psi_{i;1\alpha}+h.c.) -\Phi_a \sum_{ij }(f^\dagger_{i;\alpha}\psi_{j;a\alpha}+h.c.)
\end{align}

The mean field ansatz can be determined from the self-consistent equations:

\begin{align}
\Phi_0&=\frac{1}{4} t  N^2  \sum_{j \sim i}\chi^\psi_{ji;22} \chi^{\psi f}_{0;1} \notag\\
\Phi_a&= \frac{1}{4} J' N \chi^{\psi f}_{ji;a} \notag\\
t_f&= \frac{1}{2} J N \chi^{f}_{ji} \notag\\
t_{\psi;ab}&= \frac{1}{2} J_d N \chi^{\psi}_{ji;ba}+\frac{1}{4} t  N^2 |\chi^{\psi f}_{0;1}|^2 \delta_{a2}\delta_{b2}
 \end{align}

 where,

 \begin{align}
\chi^{\psi f}_{0;a}&= \frac{1}{N}\langle \psi^\dagger_{i;a \alpha} f_{i;\alpha} \rangle \notag\\
\chi^{\psi f}_{ji;a}&= \frac{1}{N}\langle \psi^\dagger_{j;a \alpha} f_{i;\alpha} \rangle \notag\\
\chi^{\psi}_{ji;ab}&= \frac{1}{N} \langle \psi^\dagger_{j;a \alpha} \psi_{i;b \alpha}\rangle \notag\\
\chi^{f}_{ji}&= \frac{1}{N} \langle f^\dagger_{j; \alpha} f_{i;\alpha}\rangle \notag\\
 \end{align}
In the decoupling we did not include cross terms proportional to $\delta_{\alpha \beta}$, which is smaller by a factor of $1/N$.  This should be a good approximation at least at large $N$.

\subsection{FL* phase}

\begin{figure}[ht]
\centering
\includegraphics[width=0.95\textwidth]{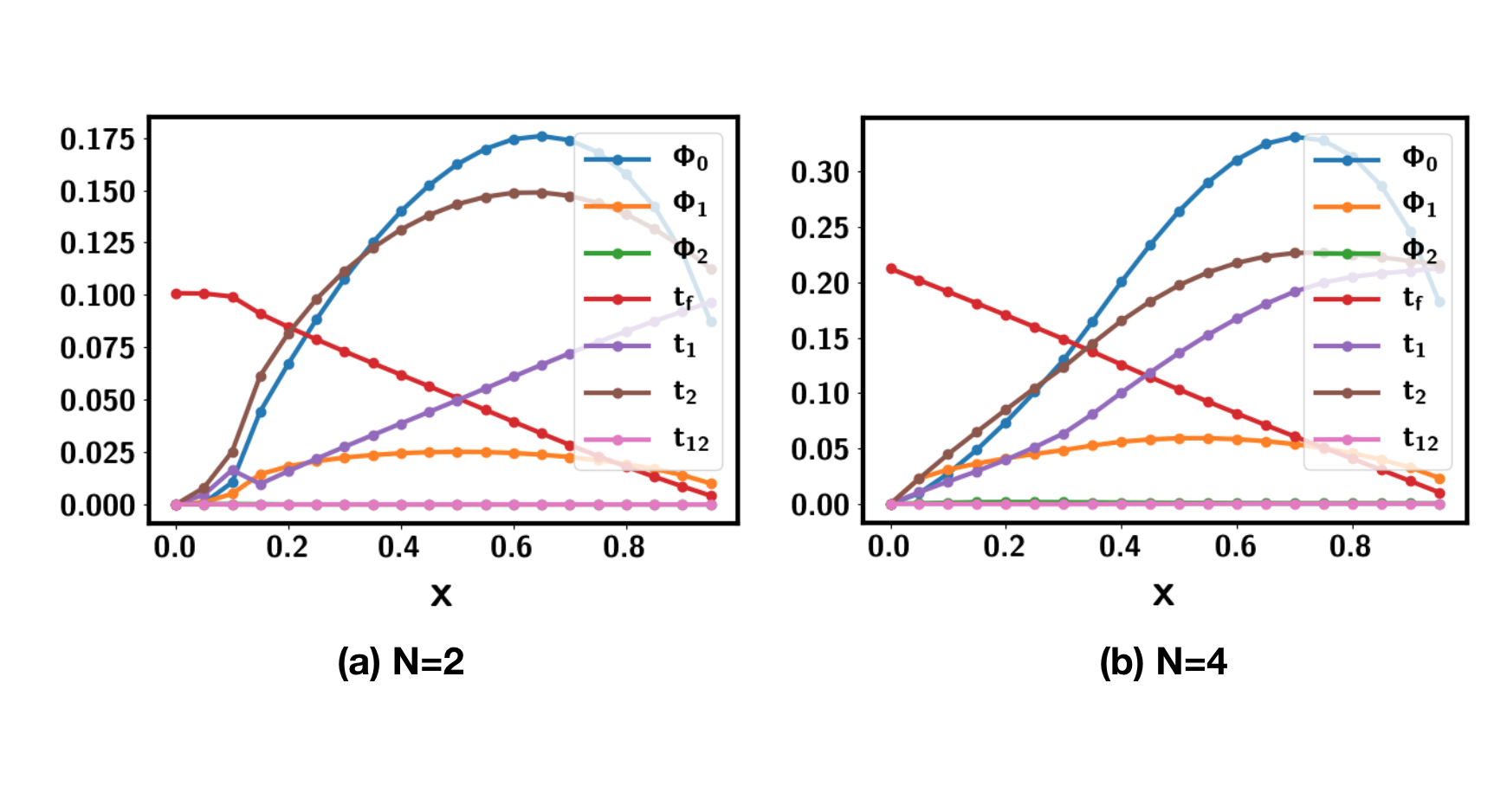}
\caption{Mean field ansatz obtained from the self-consistent equations at zero temperature with doping $x$ for $N=2$ and $N=10$ on square lattice for the proposed $SU(N)$ $t-J$ model.  We used the parameter $t=1$, $J=J_d=J'=0.5$.}
\label{fig:mean_field_append}
\end{figure}

A non-zero $\Phi_0$ and $\Phi_1$ higgs the $U(2)$ gauge symmetry down to $U(1)$.  $f^\dagger \psi_1$ couples to $a_2-\frac{1}{2}A$, hence its condensation locks $a_2=\frac{1}{2}A$.   After that, $\psi_2$ couples to $\frac{A}{2}+a_2=A$ and $f,\psi_1$ couples to $\tilde a_1=a_1+\frac{1}{2} A$.  $\tilde a_1$ remains deconfined and $f,\psi_1$ should be viewed as neutral spinons.  In contrast,$\psi_2$ couples to $A$ only and is  identical to a physical electron.    From our self-consistent calculation, $t_{\psi;12}=\Phi_2=0$ (see Fig.~\ref{fig:mean_field_append} and thus the spinon part ($f,\psi_1$) decouples from the electron part ($\psi_2$).  Therefore we have a FL* phase with a small Fermi surface coexisting with a spin liquid.  In our ansatz, the spinon forms a spinon Fermi surface.

The phase may be intuitively understood from orbital selective Mott transition.  Starting from two microscopic orbitals $d_1$ and $d_2$, we can reach the FL* phase if only $d_1$ becomes Mott localized while $d_2$ remains to form a Fermi liquid with small pocket.  However, we need to emphasize that this picture is not precise because the two orbitals $d_1,d_2$ feel an infinitely large Hund's coupling and there is no well-defined notion of microscopic orbital in our $t-J$ model.  It is better to view $\psi_1$ and $\psi_2$ as emergent orbitals.   $\psi_2$ has a finite overlap with the microscopic orbital $d_2$ only after the condensation of $\Phi_0$.    We plot the quasi-particle residue $Z$ and the effective hopping $t_{eff}=t_{\psi,22}$ of the Fermi pocket in Fig.~\ref{fig:mean_field}, one can see that  $Z$ is below $10\%$, suggesting that $\psi_2$ is not the same as the microscopic operator $d_2$.  Besides, it is heavy because the effective hopping $t_{eff}$ is an order of magnitude smaller than the microscopic hopping.

\section{More Results from  DMRG for the generalized $t-J$ model\label{append:dmrg_t_J}}

\begin{figure}[H]
\centering
\includegraphics[width=0.95\textwidth]{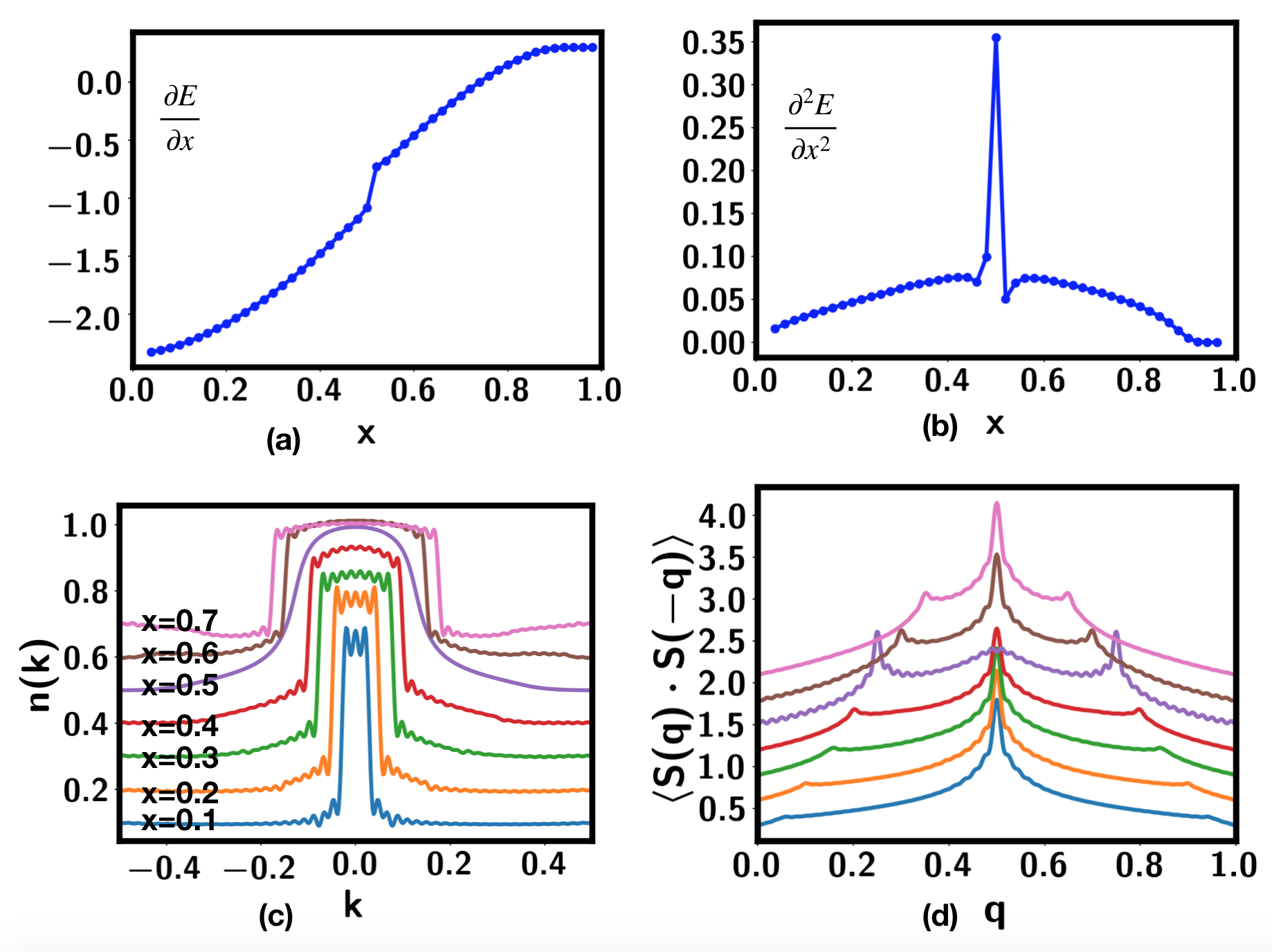}
\caption{Results from finite DMRG with $L=100$ and $D=2000$. The parameter is still $t=2J$ and $J_d=J'=J$ as defined in Eq.~\ref{eq:t_J_model_main}. Doping is varied from $0$ to $1$ with $\delta x=0.02$.  (a)(b) show a first derivative jump of energy at $x=0.5$.  In (c) and (d) we show that the system is in a LL* phase fro $x\leq 0.7$ except at $x=0.5$. At $x=0.5$ we find that $n(\mathbf k)$ does not have sharp $k_F^*$, indicating that single electron is gapped.  The fourier transformation are done using the region in $[L/4,3L/4]$ to avoid the boundary effects which still causes some wingles. Here, the momentum is in units of $2\pi$.}
\label{fig:finite_results}
\end{figure}

To change the doping with a small step, we need to use finite DMRG. Finite DMRG results of the generalized $t-J$ model with spin-one doublon are shown in Fig.~\ref{fig:finite_results}.  We find that the LL* phase is unstable at exactly $x=0.5$, at which the doublon is localized to form a charge order with momentum $Q=\frac{1}{2} 2\pi$. The LL* phase returns when $x>0.5$ and survives at least to $x=0.7$. When further increasing doping close to the spin one Haldane chain, the LL* phase may eventually be unstable to a different phase.  We do not find singularity at first and second derivative of energy when $x>0.5$, but a KT transition can not be ruled out.  We leave it to future work to study the region close to $x=1$.

\begin{figure}[H]
\centering
\includegraphics[width=0.95\textwidth]{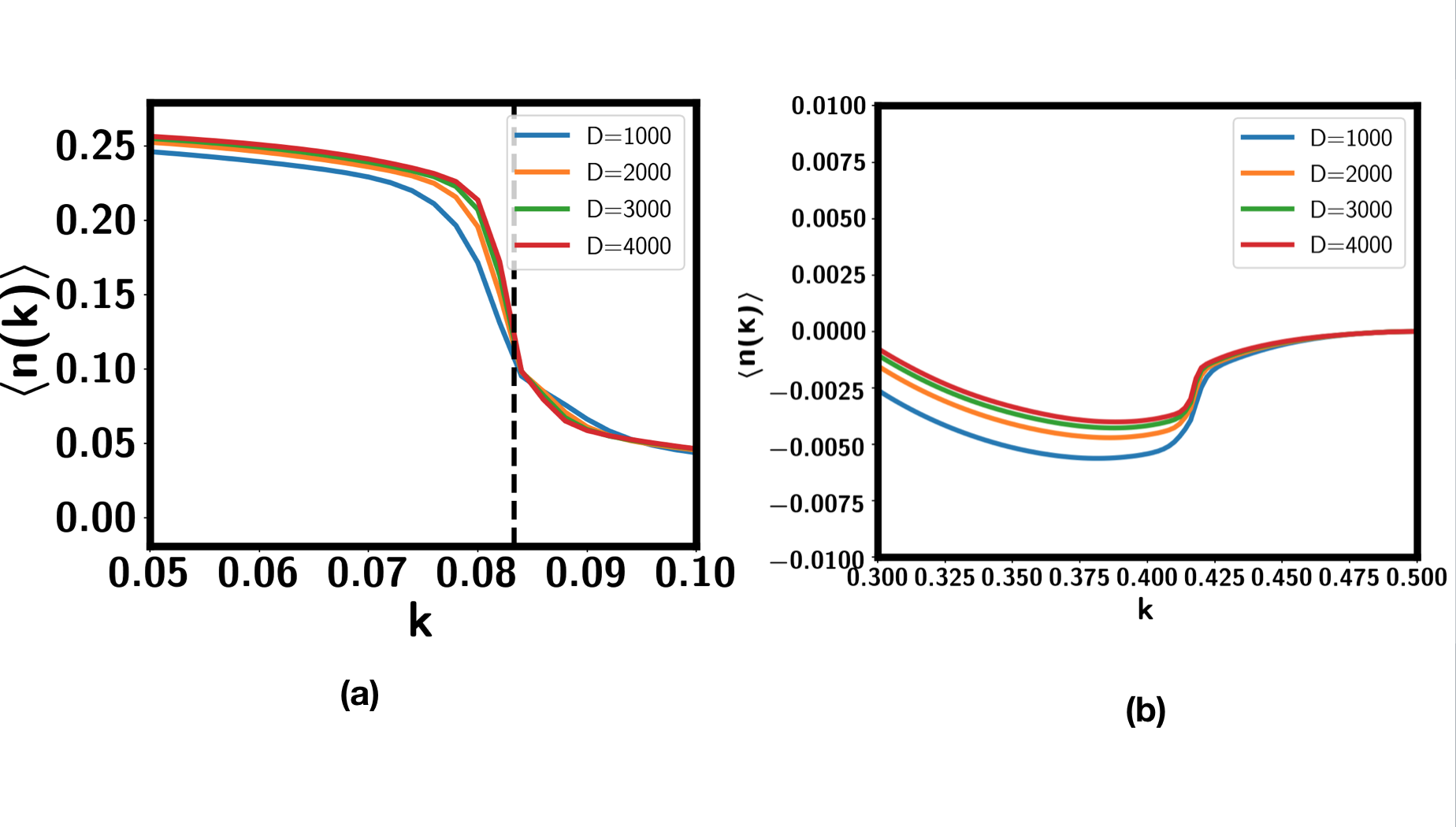}
\caption{Zoom in plots of $n(\mathbf k)$ in Fig.~\ref{fig:idmrg} of the main text for the iDMRG result at $x=\frac{1}{3}$. We focus on $k=k_F^*=\frac{x}{4}$ in Fig.(a) and $k=\frac{1}{2}-k_F^* \approx 0.416$ in Fig.(b). Here, the momentum is in unit of $2\pi$.}
\label{fig:scaling_results}
\end{figure}

As discussed in the main text, there are features at some other moemntums in addition to $k_F^*=\frac{x}{4}$ for $n(\mathbf k)$ from infinite DMRG.  In finite DMRG we do not find such features, but this may be simply because the results from finite DMRG are not as precise as in iDMRG because of the boundary effect.  To further charaterize these small features, we do a scaling with the bond dimension $D$ for the iDMRG results, as shown in Fig.~\ref{fig:scaling_results}. One can see that the rapid decrease at $k=k_F^*$ becomes steeper when we increase $D$.  In contrast, the feature around $k\approx 0.416$ actually becomes smaller at larger $D$.  This is consistent with our interpretation that the excitation here is a bound state of the elementary excitation.

%

\end{document}